# An Integrated Solar Battery based on a Charge Storing 2D Carbon Nitride

**AUTHOR LIST AND AFFILIATION**

A.Gouder[1,2], F. Podjaski[1]*, A. Jiménez-Solano[3], J. Kröger[1], Y. Wang[1], B. V. Lotsch[1,2]*

*1: Max Planck Institute for Solid State Research, Heisenbergstr. 1, 70569 Stuttgart, Germany*

*2: Ludwig-Maximilians-University, Butenandtstraße 5-13, 81377 Munich, Germany*

*3: Departamento de Física, Universidad de Córdoba, Campus de Rabanales, Edif. C2, 14071 Córdoba, Spain*

*\*: corresponding authors*

**CONTACT**

*Corresponding authors: b.lotsch@fkf.mpg.de, f.podjaski@fkf.mpg.de

## Abstract

Solar batteries capable of harvesting sunlight and storing solar energy present an attractive vista to transition our energy infrastructure into a sustainable future. We present an integrated, fully earth-abundant solar battery design based on a bifunctional (light absorbing and charge storing) carbon nitride (K-PHI) photoanode, combined with all-organic hole transfer and storage materials. An internal ladder-type hole transfer cascade via a hole transport layer is used to shuttle the photogenerated holes to the PEDOT:PSS cathode. This concept differs from previous designs such as light-assisted battery schemes or photocapacitors and allows charging with only light or light-assisted during electrical charging and discharging, thus substantially increasing the energy output of the cell. Compared to battery operation in the dark, light-assisted (dis)charging increases charge output by 243 %, thereby increasing the electric coulombic efficiency from 68.3 % in the dark to 231 %, leading to energy improvements of 94.1 % under illumination.

# 1. Introduction

While the world transitions from fossil to sustainable energy sources to tackle climate change, integrating renewable energy sources into the power grid provides its very own set of challenges. Volatile wind and photovoltaic (PV) solutions suffer from intermittent availability, which requires enhanced flexibility from both the power grid and energy storage technologies. In particular, PV produces significant stochastic intraday fluctuations (e.g., due to cloud overcast), which requires short-term energy storage solutions in the time range of minutes to hours. [1,2] Energy storage technologies can help to balance this residual load. Solar batteries and solar capacitors are a relatively new class of devices, aiming to integrate energy harvest functionalities into energy storage devices. [3,4] While discrete charging technologies (i.e., battery and PV systems as independent units) are widely employed nowadays, [5] integrated PV battery designs are attracting interest due to their more facile implementation, flexibility, and volume minimization. [6] Devices can be categorized as three-electrode configurations and two-electrode configurations: The former was demonstrated using photoelectrodes inside of batteries[7,8] or via different solar redox flow battery designs. [9-11] The latter has been enabled by depositing photoactive layers onto the anode of a battery or capacitor[12] or via composites of charge storage materials and photoactive materials. [13,14] However, a third vista is gaining momentum: Utilizing bifunctional photoelectrochemical energy storage materials, which are capable of performing both tasks at once. [15] There are several reports of materials showing photo-assisted charging, both inorganic (e.g. $V_2O_5$, [16,17] $MoO_3$, [18] $TiO_2$, [19] or 2D perovskites[20]) and organic (e.g. covalent organic frameworks, [21] quinone derivatives, [22] or porous organic cages[23]). However, most reported devices work in a light-assisted electric charging mode and are not conducive to charging by light only. Notably, all existing device designs rely on transfer of the photogenerated charge carriers to the anode or cathode via an external circuit during charging, with the separator between the electrodes acting solely to conduct ions and prevent a short circuit – as inspired by traditional battery designs – and thus resemble integrated PV-batteries. On the other hand, designing a solar battery derived from a solar cell by incorporating charge storage materials on the counter electrode has received less attention and has only been demonstrated as photo(super)capacitors so far. This was achieved by incorporating a (pseudo)capacitive charge storage material and redox shuttles to close the internal circuit. [4,24,25] While such a design allows the operation of the device as a solar cell, external wiring between anode and cathode is still required during charging. Simultaneous photocharging and discharging via a load is complicated, since the external wiring is engaged in the charging process, and external electronics are necessary to change from charging to discharging mode. These drawbacks motivate us to investigate pathways of internal photogenerated hole transfer with a separator that simultaneously acts as hole shuttle.

We recently reported a bifunctional solar battery electrode material based on the fully earth-abundant 2D carbon nitride potassium poly(heptazine imide) (K-PHI). [15] Upon light excitation (bandgap of ~2.7 eV), electron-hole pair separation, and extraction of the hole, K-PHI can "trap" photoexcited electrons up to several hours[26] and release them on demand, accompanied by a color change from yellow to blue. This combination of optoelectronic and optoionic properties, which are linked to photointercalation of $K^+$ ions[27] and electron trapping in an intercalation band within the bandgap produces (pseudo)capacitive electron storage[15,28] and has led to applications in "dark" photocatalysis, [29-33] photomemristive sensing, [28] and multifunctional light-driven microswimmers. [34,35]

Herein, we design a proof-of-concept "direct solar battery" using K-PHI as active layer, which is tasked with absorbing light and storing photoexcited electrons as well as balancing charges with intrinsic $K^+$ ion movement. Contrary to currently published designs, we do not use an external wire to transfer holes to the hole storage material (HSM) counter electrode), but rather rely on an internal ladder-type hole transfer cascade performed by a multifunctional hole transport material (HTM) – a design more reminiscent of planar heterojunction-type solar cells than of batteries (Fig. 1 (a)). Photocharging occurs internally under open circuit potential (OCP) conditions, i.e., external charge transfer via wires is required for the charging process. We first identify suitable materials, poly(9,9-dioctylfluorene-alt-benzothiadiazole) (F8BT) as HTM and poly(3,4-ethylendioxythiophene) polystyrene sulfonate (PEDOT:PSS) as HSM, and then investigate kinetics and performance of different operation modes: (i) charging via illumination only and under open circuit potential (OCP) conditions, (ii) solely electric in the dark with an external current, and finally (iii) in a light-assisted electric mode.

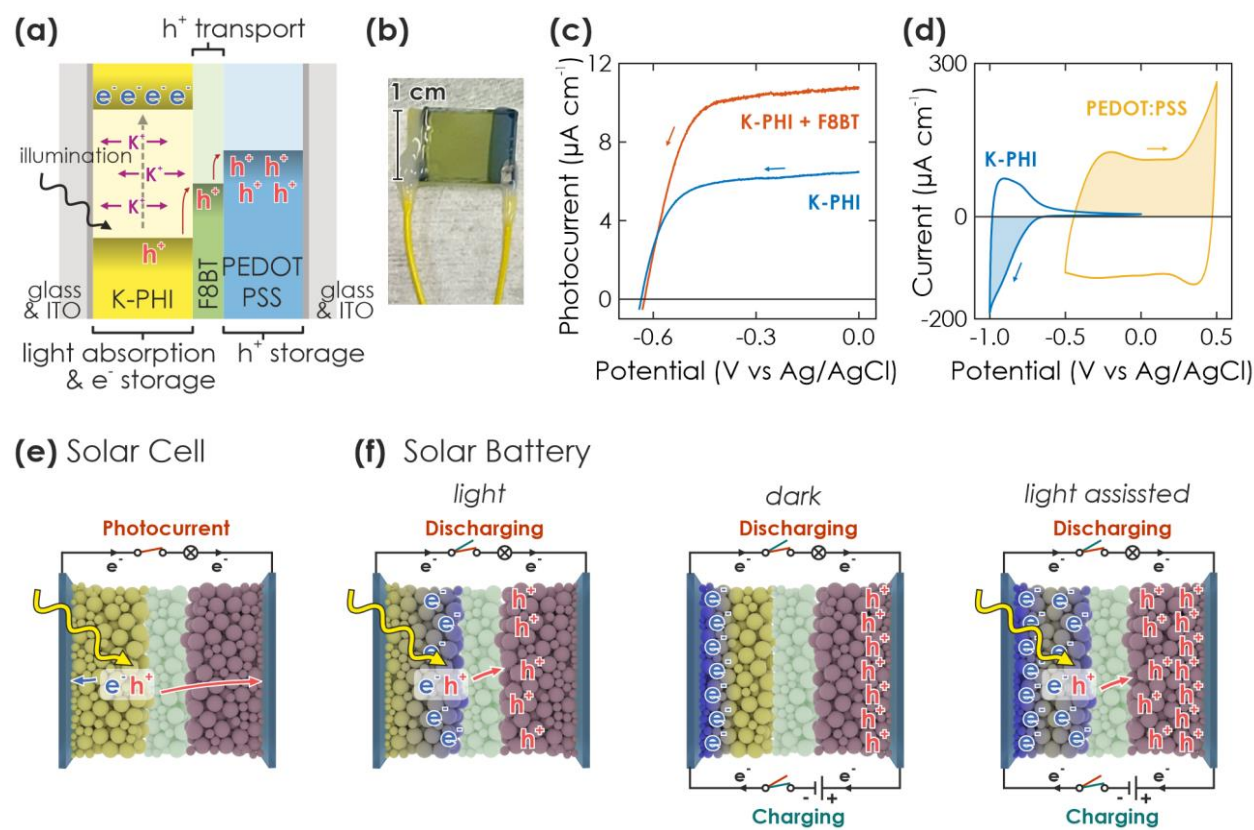

**Fig. 1 | Concept and requirements of a solar battery device. (a)** Scheme of a direct solar battery device, comprising K-PHI as photoactive and electron storage material, the HTM F8BT and the HSM PEDOT:PSS, sandwiched between two ITO sheets. The hole transport process is indicated with red arrows. **(b)** Picture of a direct solar battery device. The left and right wires are soldered to the substrate which is in contact with K-PHI and PEDOT:PSS, respectively. **(c)** Linear scanning voltammetry (LSV) curves of K-PHI (blue) and K-PHI decorated with the HTM F8BT (red), measured in an aqueous electrolyte containing the sacrificial electron donor methanol (100 mM) in 3-electrode configuration. K-PHI + F8BT shows a significantly larger photocurrent at potentials more positive than -0.4 V vs. Ag/AgCl, evidencing enhanced photogenerated hole extraction rates with the HTM. **(d)** Cyclic voltammetry (CV) measurements of ITO electrodes containing K-PHI (blue) and PEDOT:PSS (yellow), measured in 3-electrode configuration. **(e-f)** Schemes of different operation modes of the device, either as solar cell (e) or solar battery with various operation modes (f). We show the charging mechanism of the solar battery via only light (left), only electric (middle) or simultaneously using light and electric (right) power simultaneously. Color code defines the circuit switches in respective operations (red when extracting charges from the device (either as photocurrent or via discharging), green when charging the battery). Note that both circuits utilize the same connection on the device, but are plotted on top and bottom of the device to visualize different operations.

## 2. Results and discussion

### 2.1. Device design

The concept of the solar battery is visualized in Fig. 1 (a): K-PHI absorbs light and generates electron-hole pairs. Charge separation likely occurs close to the junction to the hole acceptor, [30,36] which can either be a redox shuttle or hole transport material (HTM). A solid HTM may present fewer self-discharge pathways via an electrolyte (e.g., water oxidation or reduction) and is less prone to recombination since charges are immediately shuttled to the hole storage material (HSM) – a problem which has been identified as major challenge for solar batteries. [3] In the integrated direct solar battery we propose herein, we use a solid HTM, which acts as a battery separator and redox shuttle to transport photogenerated holes from K-PHI to the HSM, while cations are shuttled between photoanode and -cathode for internal charge compensation (Fig. 1 (a), green for HTM (F8BT) and blue for HSM (PEDOT:PSS)). Simultaneously, the HTM acts as a rectifier and prevents self-discharge via an internal short-circuit between K-PHI and HSM (i.e., if holes are not only shuttled from K-PHI to the HSM, but also back from the HSM to K-PHI to quench the electrons trapped on K-PHI). The solar battery can be discharged on demand via an external electric circuit. Note that this approach with a multifunctional HTM separator is thus far unique since it does not require an external wire to shuttle charge carriers from one electrode to another during charging, further facilitating implementation by allowing simultaneous light charging and electric discharging, as well as operation as a solar cell.

We now discuss the fabrication of the device and the rationale behind the materials selection (Fig. 1 (b)): Films of K-PHI on indium tin oxide substrates (ITO) were prepared according to a procedure recently described by us. [15,28] In brief, K-PHI was synthesized in a salt melt containing KSCN and the 1D heptazine based polymer melon. [31,37,38] Subsequently, the product was washed, exfoliated via sonication in isopropanol, and homogeneous films of 0.5-2 µm thickness were obtained via dip coating (see Methods section and SI Section 1 for more details). In order to optimize charge separation at the K-PHI / HTM interface, we first performed screening experiments of both conductive polymer and small molecule HTMs (deposited via spin coating onto K-PHI) utilizing the sacrificial electron donor methanol as a replacement for the HSM (see SI Section 2 for a more detailed discussion). [15] By evaluating photocurrent as a Fig. of merit, we identified F8BT as the most suitable candidate. The IV curve of K-PHI and F8BT, measured against a reference electrode in three-electrode configuration, is shown in Fig. 1 (c) under 1 sun illumination and compared to bare K-PHI. At a potential of 0 V vs. Ag/AgCl where photogenerated electrons are discharged, an oxidative photocurrent of 10.7 µA and 6.47 µA is reached for K-PHI with and without F8BT, respectively, highlighting the beneficial role of F8BT to extract holes from K-PHI. With more negative potentials, the photocurrent decreases nearly linearly due to a decreasing driving force for electron extraction, until at -0.4 V vs. Ag/AgCl it collapses to 0 µA. For all samples, we observed an open circuit potential (OCP) of about -0.6 V vs. Ag/AgCl (Fig. 1 (c)). As HSM, we chose the widely studied conductive polymer PEDOT:PSS, deposited analogous to F8BT via spin coating. PEDOT:PSS was shown to operate as a p-type substrate capable of reductively quenching holes on n-type K-PHI upon photoexcitation. This process is akin to photocharging and underlines the suitability of PEDOT:PSS as HSM. [36] Charge storage is enabled at potentials more positive than the valence band of K-PHI and F8BT via a well investigated pseudocapacitve mechanism, making it a suitable cathode material. [39] The ladder-type redox band position (band alignment) of the HTM and HSM (Fig. 1 (a)) is thus suitable for extracting photogenerated holes in a cascade process. Three-electrode cyclic voltammetry (CV) measurements of K-PHI and PEDOT:PSS samples in the dark show the charge storage potential and capacity of both materials (see overlay in Fig. 1 (d)). While K-PHI shows its well-reported typical CV shape with a charging onset at -0.65 V vs. Ag/AgCl, [15] PEDOT:PSS produces a nearly rectangular CV – typical for its pseudocapacitve charge storage mechanism. [40] Note that the capacity of PEDOT:PSS is chosen larger than of K-PHI to prevent a performance bottleneck on the cathode side. Utilizing all these components allows us to realize the integrated solar battery (Fig. 1 (b)).

## 2.2. Operation modes of the device

A solar battery can be operated in different modes. [3,6] Upon illumination, the resulting photocurrent can be accessed by connecting the anode and cathode to the potentiostat and applying a suitable bias voltage. The device operates analogous to a solar cell (Fig. 1 (e)). However, when operating under OCP conditions (i.e., no current is extracted via the current collectors), the photogenerated electrons and holes accumulate in the anode and cathode and thus, charge the device (Fig. 1 (f), left). Subsequently, the stored charges can be accessed in the dark by applying a suitable discharge current until the cell voltage reaches 0 V (voltage at which the device is empty).

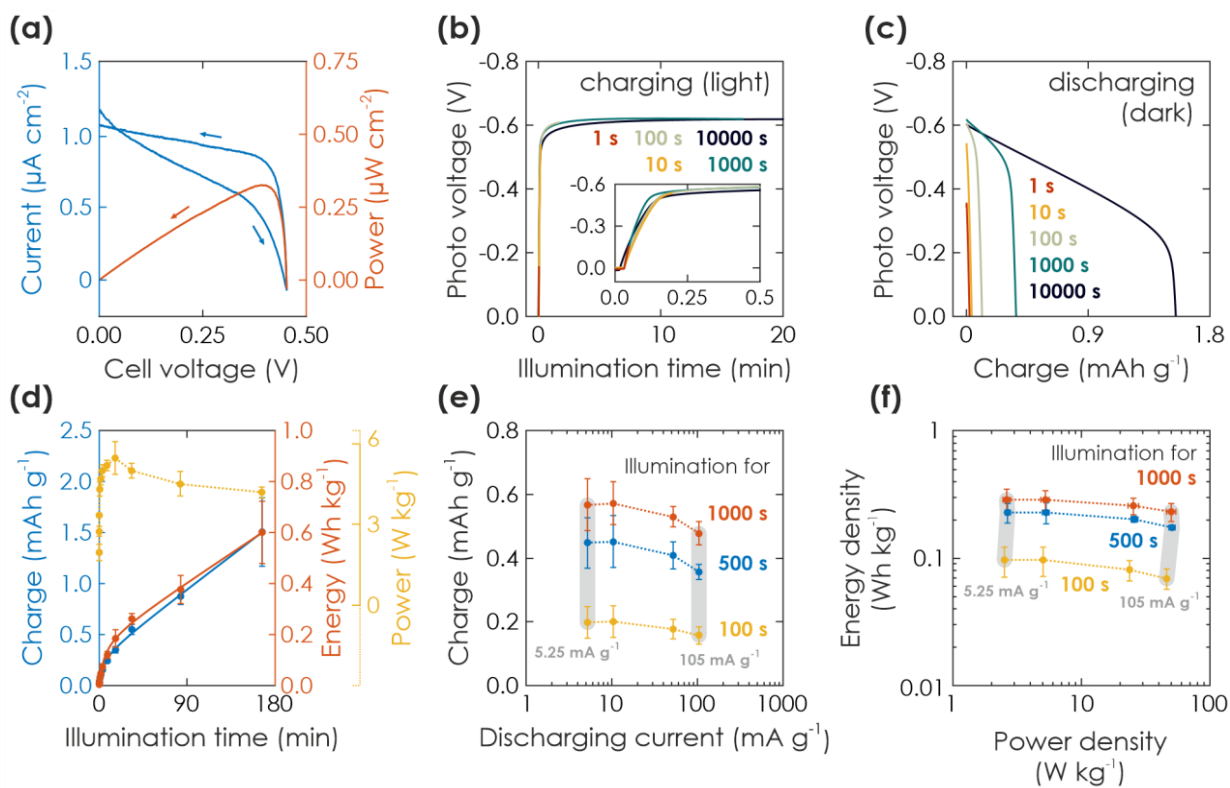

**Fig. 2 | Solar battery characterization of light charging process. (a)** Current-voltage (blue) and power (red) curves (10 mV s$^{-1}$) of a solar battery sample in solar cell mode, illuminated with 1 sun. **(b)** Charging of the solar battery with different illumination times at 1 sun and under OCP conditions (inset shows a zoom of short illumination times). **(c)** Subsequent electric discharging in the dark with a fixed current of 10.5 mA g$^{-1}$ (normalized against mass of K-PHI, HTM, and HSM). **(d)** Gravimetric capacity, energy, and power density extracted from the charging via illumination (b). **(e)** Kinetic study of the discharging process. Charging is performed via illumination for 100 s (yellow), 500 s (blue), and 1,000 s (red). Subsequent immediate discharging is carried out with different discharging currents (5.25, 10.5, 52.5, 105 mA g$^{-1}$; smallest and largest current shown with vertical grey bar). **(f)** Ragone plot displaying the energy and power output with increasing illumination times and same discharging currents given in (d). Vertical grey bars links dots measured at the same smallest and largest discharge current.

Conversely, we can also perform electric charging and discharging in the dark via an external current (i.e., a current applied via the potentiostat) – analogous to a galvanostatic charging and discharging experiment (GCD) in a normal battery (Fig. 1 (f), middle). The capacity depends on the charging and discharging current rates and voltage window. The latter should be estimated from the photovoltage measured during light charging. Notably, when illuminating the device during a GCD experiment, the current flux is created by both light generated charges and the electric charging via the potentiostat, thus maximizing performance (Fig. 1 (f), right). The overall effective charging current is the sum of both applied "external" current and "internal" photocurrent from light absorption. We will discuss these different operation modes in the following.

### 2.2.1. Solar cell operation

We first evaluate performance of the solar battery when operated as a solar cell. A device is immersed into oxygen-free 0.1 M KCl electrolyte and an initial activation measurement is performed. Activation measurements are necessary to remove all unwanted charges from both anode and cathode, which might reside on the sample from synthesis (see SI Section 3 for details) and affect device characterization. Subsequently, we illuminated the sample with an LED (365 nm, 100 mW cm$^{-2}$) and performed a CV measurement between OCP and 0 V vertex potentials with a slow scan rate of 10 mV s$^{-1}$ to simulate quasi-static conditions. The voltage sweeps are shown in Fig. 2 (a). At 0 V, we measured a short circuit current ($I_{SC}$) of 1.07 µA cm$^{-2}$ g$^{-1}$ on the backwards voltage sweep (from OCP to 0 V). Note that the mass of the device is calculated from the measured mass of K-PHI, HTM, and HSM. With increasing potential, the photocurrent is decreasing and at a potential of approximately 0.40 V it collapses to 0 V. The OCP is 0.45 V and maximum power of 0.326 µW cm$^{-2}$ is reached at a current of 0.828 µA cm$^{-2}$ and a voltage of 0.39 V, resulting in a fill factor (FF) of 0.73 (all values calculated from backwards sweep). This behavior is in principle also observed when illuminating with a 365 nm LED (100 mW cm$^{-1}$) increasing the photon flux that can be absorbed, albeit at higher absolute currents (FF of 0.70, see SI Section 9). While this FF is considered to be high for organic solar cell devices and is larger than common solar cells incorporating carbon nitrides as dyes,[41,42] the losses result from the small but significant slope of the current between 0 V and 0.30 V. This slope is probably caused by high series and low shunt resistance (e.g., due to pinholes, or traps[43]) as well as a decreasing driving force for charge separation. In case of a solar battery, current increase or loss due to a partial charging of K-PHI and PEDOT:PSS seems also possible (vide infra). The hysteresis between positive and negative voltage sweeps further indicates such a behavior: The larger current on the backward voltage sweep (i.e., from OCP to 0 V) might be a convolution of photocurrent and discharging current of K-PHI. Such a behavior is desirable for the solar battery light charging modes discussed next.

### 2.2.2. Solar battery operation *via light charging*

We can also employ the internal photocurrent to charge the device in lieu of extracting it immediately as discussed in the previous section. This enables the most characteristic function of a solar battery: its ability to charge solely via illumination. We demonstrate in an experiment charging under illumination and electric discharging in the dark in Fig. 2 (b) and (c): After immersing a solar battery sample into a degassed 0.1 M aqueous KCl electrolyte and performing the activation measurement (see SI Section 3), we illuminated the device from the backside for a given time at 1 sun under OCP conditions. Note that the task of the 0.1 M KCl electrolyte in the reactor is only to provide an oxygen-free environment, ensure stable temperature during illumination, provide sufficient humidity and to facilitate reset measurements (see SI Section 3). A photovoltage of 0.6 V developed during the first 50 s and remained constant during the ensuing illumination. Subsequently, the light was turned off and the device was discharged at a current of 10.5 mA g$^{-1}$ until the cell voltage dropped back to 0 V. Electric discharging in the dark for increasing illumination charging times (1s to 10,000 s) is shown in Fig. 2 (c). The shape looks similar to a typical GCD battery measurement and indicates a faradaic charge storage mechanism: A plateau-like potential decrease to ca. 0.5 V for short and 0.3 V for long illumination times, followed by a sharp voltage drop to 0 V. Note that the plateau has a certain slope < 0, indicating pseudocapacitive contributions of the charge storage mechanism, and possibly also caused by increasing PHI or cell resistance upon discharging.[15,28,40] We discuss the charge storage mechanism more thoroughly via a kinetic analysis with CV measurements in SI Section 4.

The respective capacity, energy, and power output is plotted against the illumination time in Fig. 2 (d). Energy output can be calculated with E = ∫ I$_{ph}$·V(t) dt, with I$_{ph}$ being discharge current and V(t) being the cell voltage during discharging. Average power is calculated with P = E·t$^{-1}$, with t being the discharging time. When illuminating the sample for 10,000 s, we can extract a charge of 1.5 mAh g$^{-1}$ and an energy of 0.60 Wh kg$^{-1}$. A saturation behavior of the capacity becomes evident for illumination times above ca. 2,000 s (Fig. 2 (d), blue: initial slope is decreasing), but is not reached after 10,000 s, suggesting that the charging process is likely slowed down by the concomitant charge accumulation[28,44,45] limiting photocharging efficiency and that the true capacity of K-PHI is hence larger. Noteworthy, the shape of the extracted charge and energy curve is similar, since energy output depends on the photovoltage V(t), which is almost constant at illumination times where the plateau is reached (> 100 s; see Fig. 2 (d), yellow). Power output for these illumination times is approximately constant since it is governed by the photovoltage as well. We discuss self-discharge in SI Section 8 and show a charge retention of 72 % after 1000 s, when the device is illuminated for 1000 s.

Next, in order to analyze discharging kinetics, a solar battery sample was illuminated with 1 sun for three representative durations (100, 500, 1,000 s) under OCP conditions and subsequently discharged in the dark with different current densities (5.25, 10.5, 52.5, 105 mA g$^{-1}$). The extracted charge is shown in Fig. 2 (e). A smaller discharging current results in a larger capacity as common for batteries due to less diffusion limitations and less resistive losses resulting from the intrinsically low conductivity of K-PHI.[15,46] While we observe this larger capacity with lower currents for all illumination times, it is more pronounced for longer durations (when comparing currents of 5.25 and 105 mA g$^{-1}$: 26 % and 19 % larger capacity for the smaller discharge current at illumination times of 100 s and 1,000 s, respectively). Scaling of energy and power density with current is presented via a Ragone plot in Fig. 2 (f): Energy density behaves analogous to the capacity (SI Fig. 6.1), i.e., a minor increase is observed for smaller currents (5.25 - 10.5 mA g$^{-1}$). With illumination duration the energy density increases for all currents (Fig. 2 (f): yellow to red) – analogous to the behavior of energy and charge scaling with illumination time discussed above. Power density increases with current, since the cell voltage for different discharge currents is approximately constant (see GCD profiles in SI Fig. 5.1 and respective power output in SI Fig. 6.1), in line with the discussion of power output dependence on illumination time above. Thus, the kinetic behavior results in a Ragone plot with rather horizontal lines, which scales upwards (i.e., increase energy density) with illumination time, highlighting the device's energy-stable operation at larger discharge currents, i.e., increased power density (up to approx. 10 W cm$^{-2}$ kg$^{-1}$). We will discuss comparison to energy storage and solar battery devices in more detail in Section 2.2.3 in the context of light-assisted electric charging and discharging.

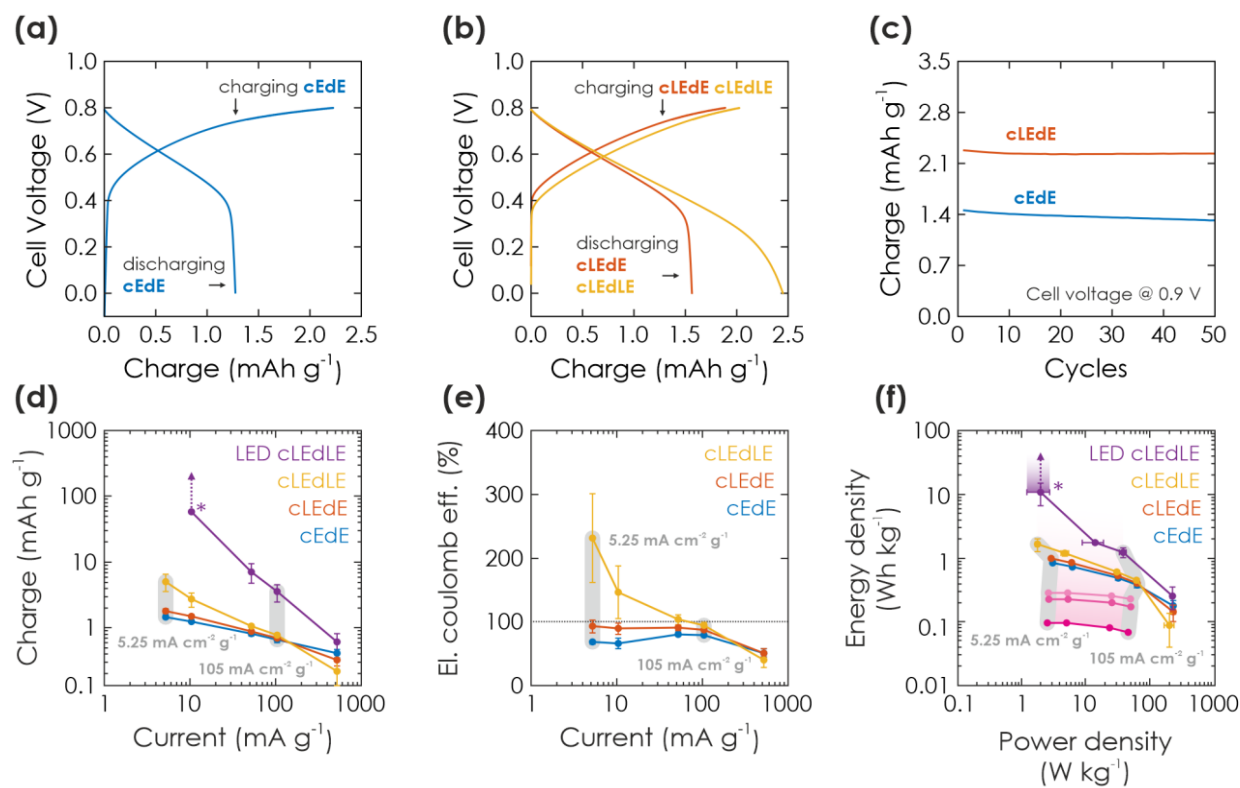

**Fig. 3 | Solar battery characterization of electric and light-assisted electric charging process. (a-b)** A cycle showing GCD in the dark (a) (cEdE) and under 1 sun illumination (b) during charging (cLEdE) or during charging and discharging (cLEdLE) with a current of 10.5 mA g$^{-1}$. **(c)** Extracted charge in cEdE and cLEdE mode and a cell voltage window of 0.9 V, plotted against the cycle number. **(d)** Extracted charge as a function of electric charging and discharging current, when operating the solar battery in a voltage window of 0.8 V and cEdE, cLEdE or cLEdLE modes. Note that LED-cLEdLE is analogous to cLEdLE, but with a 365 nm LED (power analogous to 1 sun: 100 mW cm$^{-2}$) providing the illumination. The "*" marks a data point, where discharging could not be completed and the measurement was aborted (due to the photocurrent being larger than the discharging current). The vertical arrows show how charge would have increased with a later aborted measurement. The two exemplary vertical gray bars show measurements at the same charging and discharging current. **(e)** Electric coulombic efficiency for the measurement shown in (d), highlighting the performance gains from illumination. **(f)** Ragone plot, showing energy and power output at the currents given in (d) and (e), underlining how illumination can push the device performance to larger energy and power values. Charging solely via illumination is extracted from Fig. 2 (e) and shown in purple, with illumination times increasing from 100 s to 1,000 s from dark to bright purple. Arrow and "*" mark again the aborted measurement (as discussed in (d)).

### 2.2.3. Solar Battery operation *via electric and light-assisted electric charging*

So far, we have discussed the ability of the solar battery to charge via illumination. However, as a second pathway to modify the charging state, analogous to a normal battery we can apply an external electric charging current, which can be combined with the internal photocurrent to store both solar and electric energy simultaneously in one device. To differentiate between different illumination modes during GCD, we use the terminology "cEdE" to describe electric charging and discharging in the dark, "cLEdE" when illuminating during electric charging, and "cLEdLE" when illuminating during both charging and discharging. Cell efficiency metrics defined by parameters such as capacity, energy and power density, or electric coulombic efficiency (eCE) can significantly improve by these combined methods. The eCE gives the ratio between charges being electrically discharged and electrically charged and thus, presents a metric to evaluate charge gains via illumination (for cEdE: eCE is analogous to the coulombic efficiency) In the following, we will discuss solar battery operation modes, which include electric charging via an applied current to modify the charging state.

We first perform GCD with a current of 10.5 mA g$^{-1}$ from 0 to 0.8 V in similar conditions as for the light charging measurements discussed above (see SI Section 7 for the rationale behind the chosen voltage window), but in the dark (cEdE mode). We show a cycle in Fig. 3 (a). For charging, we observe an initial fast voltage increase to about 0.5 V, followed by an area with significantly slower voltage increase (i.e., "plateau" region). When discharging, we observe a similar trend: slow voltage decrease from 0.8 V to 0.5 V and a subsequent collapse to 0 V. Note that the shape of the discharge curve looks similar to when the sample is only charged via illumination for short illumination times (Fig. 2 (c); albeit at ca. 0.2 V smaller voltage), hinting onto a similar discharging mechanism.

Light-assisted GCD can be performed in cLEdE (Fig. 3 (b), red) and cLEdLE (Fig. 3 (b), yellow) operation modes. Compared to the dark measurement (Fig. 3 (a)), for charging we observe again an initial region of fast voltage increase to about 0.5 V and a subsequent "plateau" region with a slope comparable to the dark measurement. However, for discharging the "plateau" lasts longer to about 0.4 V for cLEdE and 0.2 V for cLEdLE, analogous to charging with long illumination times (Fig. 2 (c)). This results in more extracted charge, which we show in Fig. 3 (c) for a cell voltage of 0.9 V: Capacities of both dark and light GCD measurements are compared for 50 cycles, yielding an initial capacity of 1.4 mAh g$^{-1}$ and 2.3 mAh g$^{-1}$ (92 % and 98 % retained after 50 cycles), respectively. To explain mechanistic differences in electric charging via GCD (discussed here) and purely light charging (discussed in Section 2.2.2), we assume that GCD charges K-PHI and PEDOT:PSS close to the current collector (ITO substrate) first. In contrast, photocharging requires interfacial charge separation and therefore occurs rather at the junction to the HTM, or at least more in the bulk of the active layer of the battery (visualized in Fig. 1 (f): compare position of electrons in left vs. middle panel). Since photogenerated electrons close to the HTM junction have to travel through the bulk of the K-PHI layer to discharge via the substrate, charge transport limitations due to the internal resistance of K-PHI (i.e., larger iR drop) will have a larger effect as compared to electrons injected close to the substrate. Thus, self-discharge for light charging must be larger than for electric charging, which we show and discuss in SI Section 8. Besides affecting discharging kinetics, the larger iR drop for photocharged electrons reduces the final cell voltage (compare cell voltage at the end of the discharging plateau in cEdE (0.5 V, Fig. 3 (a)), cLEdE and cLEdLE (0.4 V and 0.2 V, Fig. 3 (b))). Thus, simultaneous charging via light and the substrate allows to access more of the solar battery volume on short time scales, which benefits the overall capacity and thus energy density. In addition, illumination increases the material's conductivity, [15] thereby facilitating also the electric charging process.

To deepen our understanding of the influence of light on parallel electric charging, we perform a kinetic study by changing charging and discharging currents. In Fig. 3 (d), we show the respective charge output for cEdE, cLEdE, and cLEdLE modes (LED-cLEdLE is analogous to cLEdLE, but uses a 365 nm LED (100 mW cm$^{-2}$) as light source). Note that measurements were performed in a smaller voltage window of 0.8 V compared to 0.9 V used for the cycling stability discussed above (Fig. 3 (c)) due to the more stable operation in this voltage window at small currents and under illumination, allowing for a more reliable comparison of different operation modes. When operating in cEdE mode, with smaller currents (5.25 – 10.5 mA g$^{-1}$) the capacity increases as the system is kinetically less limited. At the same time, this increase starts to saturate for very small currents due to the longer discharging time invoking self-discharge (a discussion on self-discharge after electrical charging is given in SI Section 8). When operating in cLEdE mode (Fig. 3 (d), red), the change of capacity behaves analogous to the dark case, but with an offset to larger capacities (compared to the dark case: at a current of 5.25 mA g$^{-1}$ we observe an increase of extracted charge of 22.0 % to 1.79 mAh g$^{-1}$). The offset results from the internal photocurrent assisting charging the device as discussed above. This effect is more pronounced for smaller currents (at a current of 105 mA g$^{-1}$ we could only observe an increase of 5.54 %), which we explain with the longer charging time resulting from small currents leading to an elongated illumination time. When operating in cLEdLE mode (Fig. 3 (d), yellow), the overall illumination time becomes much longer since the internal photocurrent is also continuously generated during discharge, which in return significantly increases the extracted charge output (compared to the dark case: at a current of 5.25 mA cm$^{-2}$ g$^{-1}$ we have observed an increase of extracted charge of 243 % to 5.02 mAh g$^{-1}$). Note that when the internal photocurrent is in the range or larger than the external discharging current, the extracted charge increases very significantly (see cLEdLE in Fig. 3 (d) for small currents) or even rises into infinity. We demonstrate the latter by providing illumination via a LED at similar illumination power compared to solar simulators with 1 sun (100 mW cm$^{-2}$), but which only illuminates at wavelengths of ca. 360-375 nm where K-PHI can absorb (Fig. 3 (d), LED-cLEdLE). IV curves of illumination via 1 sun (Fig. 2 (a)) and 365 nm LED are compared in SI Section 9. At a discharge current as low as 10.5 mA g$^{-1}$, we could never complete discharging due to strong continuous internal photocurrent generation and had to abort the measurement after extracting a charge of 20 mC (58.4 mAh g$^{-1}$). Hence, the observed increase in extracted charge is larger than for reported devices with a similar bifunctional electrode, but different device designs (57 % and 95 % increase in capacity for $V_2O_5$ photocathodes for lithium- and zinc-ion batteries[16,17]). This increase in charge should not be confused with an increase of capacity of the battery, but rather demonstrate a beneficial operation mode of a solar battery due to continuous charge generation under illumination.

Next, we compare the eCE for the different solar battery operation modes (Fig. 3 (e)), a metric which we define as the ratio between electric charging and electric discharging. Note that in comparison to the coulombic efficiency (CE), eCE can exceed 100% since charge generation stemming from the internal photocurrent is not taken into account in the electric external charging current.[21,22] In cEdE mode, we reach the maximum eCE (here the same as CE) of 80.7 % at a current of 52.5 mA g$^{-1}$. Larger as well as smaller currents produce a decreased eCE due to kinetic limitations and self-discharge, respectively.[15] The eCE value is in fact larger than for K-PHI in a half-cell configuration reported by us earlier (approx. 72 %[15]), which we explain with less self-discharge via the inevitable aqueous electrolyte of half-cell measurements (enabling water reduction via uncovered parts of the substrate). When operating in cLEdE, we can alleviate the eCE for small currents since additional charging occurs via the photocurrent. Thus, this mode allows a more efficient operation of the solar battery in a region where the capacity is larger (i.e., smaller currents), with a maximum eCE of 92.9 % (for cEdE: 68.3 %) at a current of 5.25 mA g$^{-1}$. When operating in cLEdLE, we see a significant increase in eCE for small currents, with a maximum of 231 % at a current of 5.25 mA g$^{-1}$. Analogous to our rationale behind the increase of extracted charge discussed above, we explain this behavior with a significantly longer illumination time compared to cLEdE, resulting in much more photocharging. Thus, the eCE reported here for cLEdLE is much larger than for literature reports of 112 %.[22] Note that eCE is not a cell efficiency, since it only takes into account electron flux (i.e., current) into and out of the cell and does not include the incoming photon flux, which is responsible for the photocurrents and hence, additional charges measured.

Finally, we compare power and energy density for different currents via a Ragone plot (Fig. 3 (f)). The cEdE measurement resembles the behavior of a normal battery:[40] Maximum energy density (0.846 Wh kg$^{-1}$) for small power at small currents and maximum power density (62.3 W kg$^{-1}$) for small energies at large currents. This behavior is caused by kinetic limitations of the discharging process, as discussed above. When operating in cLEdE or cLEdLE mode, we observe a similar curve shape as the dark case, but with an offset to larger energy densities (compared to cEdE at maximum energy: 16.0 % and 94.1 % increase to 0.982 Wh kg$^{-1}$ and 1.64 Wh kg$^{-1}$, respectively). This effect is more pronounced for smaller currents and can be explained analogous to our abovementioned rationale for larger capacities and improved eCE with internal photocurrent assisting in charging of the bulk. Power density slightly decreases concurrently due to the altered GCD discharging profile (compare Fig. 3 (a) and (b)). Increasing illumination intensity via an LED (ca. 360-375 nm, 100 mW cm$^{-1}$) significantly amplifies the energy enhancement: Simultaneous to our discussion on extracted charge, we had to abort the measurement after extracting 10.8 Wh kg$^{-1}$ since the discharging process could not be completed, but upon longer operation the energy output should approach infinity, as indicated by the purple arrow in Fig. 3 (f).

To understand the origin of performance improvements in the Ragone plot for cLEdE, cLEdLE, and LED-cLEdLE mode, we show the performance of charging solely via illumination as discussed in Section 2.2.2 in the same plot (Fig. 3 (f), purple): Longer illumination times lead to more photocharging, i.e., vertical scaling in the Ragon plot. Illumination assisted GCD measurements discussed here are affected in an analogous manner, i.e., energy increases with illumination time. Power density on the other hand scales with current and is more or less independent of illumination duration, when not taking the effect of altered GCD shapes (Fig. 3 (b)) into account. Thus, its scaling in the Ragone plot looks similar for all operation modes including charging solely via illumination (compare Fig. 3 (f) data points at different currents, marked with gray bars). A direct comparison of energy and power scaling with discharge current for all operation modes shows the similar scaling best and is given in SI Fig. 6.1.

## 3. Conclusion

In this work, we have presented a proof-of-concept integrated solar battery device based on the earth-abundant carbon nitride K-PHI, which serves as both light absorber and charge storage (photo)anode. Photogenerated holes in K-PHI are shuttled via a HTM F8BT to the HSM PEDOT-PSS via an interfacial hole transfer cascade. The device can work as a solar cell; however, its capabilities exceed those of a solar cell: When kept under OCP (i.e., no current is applied) the generated internal photocurrent charges the photoactive material and the HSM. No external wiring is required for this charging process. Subsequently, the charge can be accessed by applying a suitable discharge current. We also demonstrate purely electric charging via a charging current as a second path of accumulating charge on the device and also combine both modes, resulting in light-assisted electric (dis)charging that enables to boost the device performance further. Via kinetic studies, the performance limitations and metrics of the device are discussed while showing how Ragone plots can be used and behave for such a devices. We summarize performance parameters for different modes and kinetics in Fig. 4.

An important message of this work is to provide a fundamental understanding of the solar battery operations as a convolution of different ingoing or outgoing energy fluxes, which impact the charging state. The charging contributions are as follows: (i) Energy input via illumination depends on the generated internal photocurrent, which itself relies on the material's absorption profile, incident photon flux and illumination time (Fig. 4 (a)). (ii) Energy input via electric charging in the dark (Fig. 4 (b)) emulates a classic battery (cEdE), i.e. the capacity increases with lower charging and discharging currents due to smaller kinetic limitations. We show that a combination of light and electric charging during either charging (cLEdE) or both charging and discharging (cLEdLE) yields a performance enhancement in terms of apparent discharge energy, apparent capacity, and eCE. Thus, we provide an energy storage device, the apparent performance output of which can be tuned via illumination (see Ragone plot in Fig. 3 (f)) and relies on a double functionality of light absorption and electron storage in a single material. The device design based on a ladder-type internal hole transfer cascade, which so far is the first of its kind for solar batteries, renders it a closer relative to solar cells than to classic batteries, thus establishing a new generation of direct solar batteries derived from solar cells with bifunctional (both light absorbing and charge storing) components. The presented integrated design based entirely on earth-abundant electrode materials makes production and application facile (i.e., no external wiring is required for the charging process, simultaneous light charging and electric discharging is possible) and underscores the application potential of this new concept of truly integrated solar batteries based on bifunctional materials, especially where low cost and high levels of integration are key, for example in autonomous microsystems, self-powered sensors, or solar battery parks.

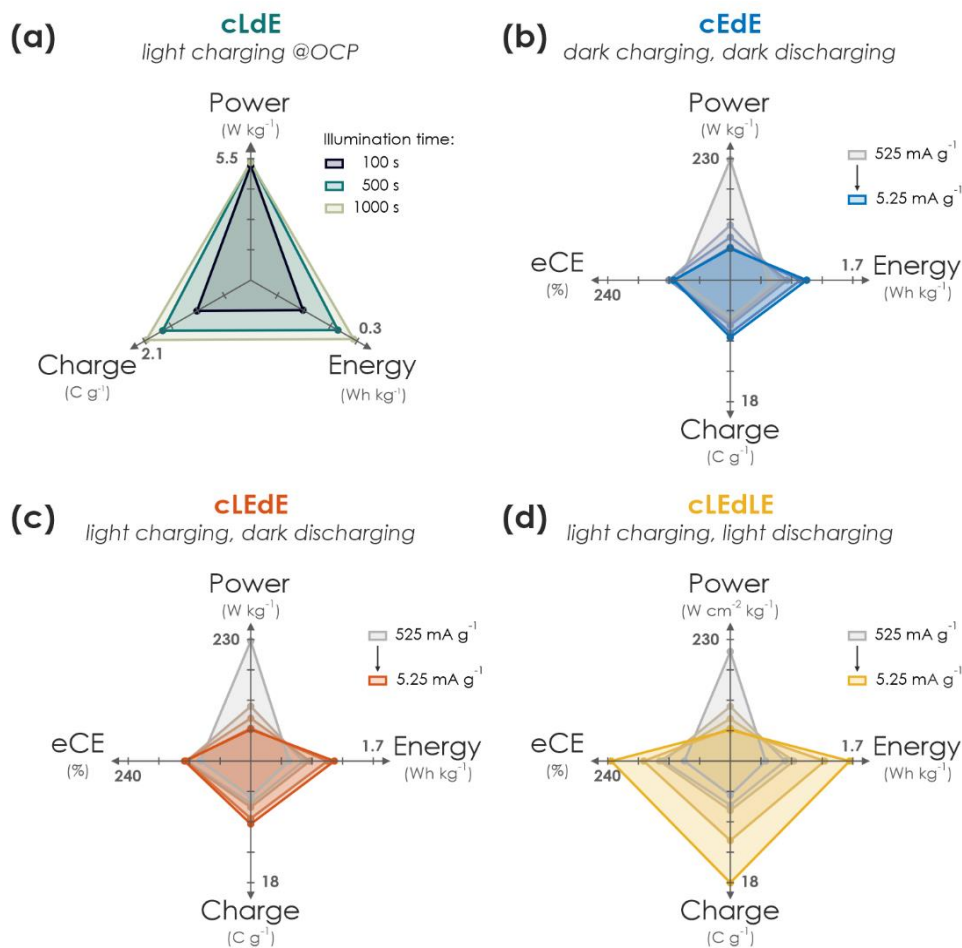

**Fig. 4 | Performance summary of operation modes of the solar battery. (a)** Pure photocharging at OCP and discharging in the dark (cLdE) at a constant current of 10.5 mA/g. While power output remains approximately constant, charge and energy

scales with illumination time. **(b)** GCD in the dark (cEdE) with different current densities (current decreases from grey to blue). With decreasing currents, power decreases and energy, charge as well as eCE slightly increases. **(c)** Same plot as in (b) for electric charging with different currents, but under illumination (1 sun) during charging for additional photocharging (cLEdE). Discharging is performed with the same current in the dark. Power, energy, charge, and eCE scale with current akin to (b), but with slightly enhanced performance. **(d)** Same plot as in (b) and (c), but under illumination during both charging and discharging (cLEdLE). While power scales analogous to (b) and (c) with current, energy, charge and eCE is significantly enhanced thanks to the solar cell output during both charging and discharging, increasing the device performance significantly compared to when operated in the dark.

## 4. Methods

### 4.1. Synthesis of K-PHI & anode preparation

K-PHI was synthesized as described in literature via a salt-melt of potassium thiocyanate (KSCN) and melon. [29,37,38] Precursors KSCN and melamine were acquired from Sigma Aldrich in reagent grade purity. Exfoliation was performed via sonication in an ice bath for 2 h in a 2-propanol (IPA) suspension (*Sigma Aldrich*) with a concentration of 3 mg/mL. Nanosheets were separated via two consecutive centrifugation steps (353 RCF for 20 min, then 795 RCF for 40 min; 3-30k, *Sigma*), akin to reported procedures. [15,28] Density was evaluated by drying 1 mL of suspension and subsequent weight of the residue and particle concentration was increased to 0.3 mg/mL by evaporating the respective amount of IPA via a rotary evaporator. Films were deposited onto the transparent conductive substrate indium tin oxide (ITO; *Ossila Ltd.*) via dip coating (ca. 400 dips, 100 mm/min extraction speed, 2 min drying between dips; Rotary dip coater, *Nadetech*), yielding films of approx. 500 to 1,000 nm.

Subsequently, for depositing the HTM a solution of F8BT (*Ossila Ltd.*) in chloroform with a concentration of 10 mg/mL was spin coated (2,000 rpm for 30 sec; WS-650MZ-23NPP, *Laurell*) onto the K-PHI film and annealed for 10 min on a hot plate at 80 °C. This process was repeated to increase the HTM film thickness. The sample was contacted by scratching off a small part of both films at a corner of the sample to uncover the substrate and by gluing a wire to it with conductive silver paste (Silver Conductive RS 186-3600, *RS-Pro*). The contact was then sealed with epoxy glue (DP410, *3M Scotch-Weld*) to provide both a rigid connection and prevent the silver paste as well as ITO to participate in the measurements.

### 4.2. Cathode preparation

For the cathode, an aqueous suspension of PEDOT:PSS with a concentration of 3-4 wt% (*Sigma Aldrich*) was spin coated (2,000 rpm, 30 sec; WS-650MZ-23NPP, *Laurell*) onto an ITO substrate (*Ossila Ltd.*) with dimensions of 10x12 mm, which underwent plasma cleaning with oxygen plasma (Femto, *Diener*) for 10 min prior to deposition to both clean the surface and make it more hydrophilic. Subsequently, the sample was annealed for 20 min at 145 °C in a nitrogen atmosphere. This process was repeated to increase PEDOT:PSS film thickness to approx. 600 nm. The samples was then contacted analogous to the anode as described in the previous section.

### 4.3. Fabrication of the solar battery full cell

First, adsorbed oxygen was removed from both anode and cathode half-cell samples by applying vacuum and subsequent argon for 6 cycles using Schlenk techniques. Both samples were then sandwiched onto each other, a weight of approx. 15 g was put on top to provide enough pressure, and epoxy (DP410, *3M Scotch-Weld*) was applied on the two opposite edges to generate a sturdy connection, while leaving the two other faces open to enable contact with the surrounding electrolyte. Subsequently, after drying of the epoxy the sample was immersed in a degassed aqueous 0.1 M KCl solution. The mass of the device was calculated by weighting the K-PHI / HTM sample after film deposition on the substrate and calculating the mass of the PEDOT:PSS sample from measured film thickness and density of dried films reported by the manufacturer.

### 4.4. (Photo)electrochemical measurements

All electrochemical measurements were performed in a photo(electrochemical) reactor equipped with a quartz glass for illumination of the respective sample, and with a multichannel potentiostate (Autolab M204, *Metrohm*). An aqueous solution containing 0.1 M potassium chloride (KCl; *Sigma Aldrich*) was used as background electrolyte (for experiments requiring a sacrificial electron donor, the respective donor was added as well), which was degassed with argon (> 99 %) prior to every measurement to ensure oxygen free conditions. For three-electrode-measurements, we have utilized an Ag/AgCl reference electrode with a saturated KCl electrolyte (RE-1CP, *ALS Japan*) and a gold foil (*Sigma Aldrich*) counter electrode. For two-electrode-measurements, the sample was directly connected to the potentiostate. Illumination was provided either with a calibrated solar simulator (LightLine A4, *Sciencetech*), providing artificial sunlight according to AM 1.5G with class AAA quality, or with a LED at 365 nm equipped with a collimator (M365LP1-C4, *ThorLabs*) and set to a power output of 100 mW cm$^{-2}$.

## 5. Acknowledgements

The authors acknowledge N. Vargas-Barbosa and L. Yao for fruitful discussions. Financial support is gratefully acknowledged by the Max Planck Society, the European Research Council (ERC) under the European Union's Horizon 2020 Research and Innovation Program (grant agreement no. 639233, COFLeaf), the Deutsche Forschungsgemeinschaft (DFG) via the Cluster of Excellence 'e-conversion' (project number EXC2089/1–390776260) and by the Center for NanoScience (CENS). A.J.S. gratefully acknowledges a postdoctoral scholarship from the Max Planck Society.

## 6. Author contributions

AG, FP, AJS, and BVL conceived the project. AG performed the measurements. AG and FP, with assistance of YW, analyzed the data. FP, AJS, and BVL supervised the research. AG and FP wrote the manuscript with assistance of all authors.

## 7. Declaration of interests

The authors declare no conflict of interest.

# Supplementary Information

An Integrated Solar Battery based on a Charge Storing 2D Carbon Nitride

**AUTHOR LIST AND AFFILITATION**

A.Gouder[1,2], F. Podjaski[1]*, A. Jiménez-Solano[3], J. Kröger[1], Y. Wang[1], B. V. Lotsch[1,2]*

*1: Max Planck Institute for Solid State Research, Heisenbergstr. 1, 70569 Stuttgart, Germany*

*2: Ludwig-Maximilians-University, Butenandtstraße 5-13, 81377 Munich, Germany*

*3: Departamento de Física, Universidad de Córdoba, Campus de Rabanales, Edif. C2, 14071 Córdoba, Spain*

*\*: corresponding authors:* b.lotsch@fkf.mpg.de, f.podjaski@fkf.mpg.de

## Table of content

# 1. Morphological characterization

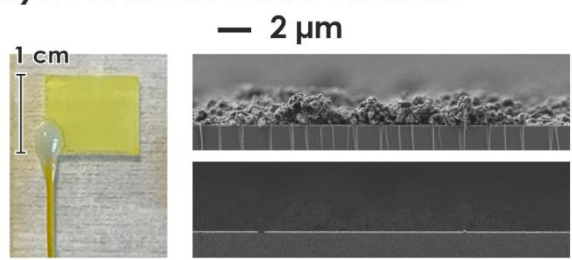
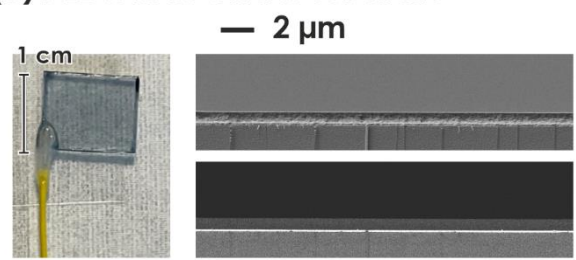
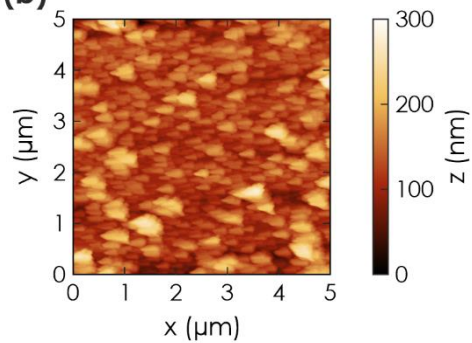
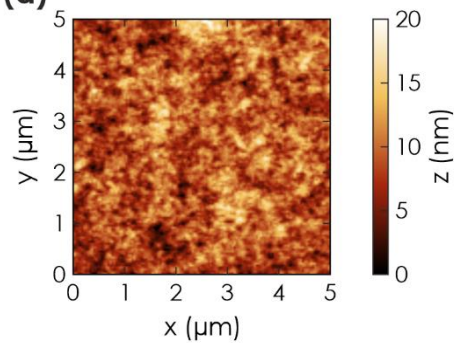

**Fig. S1.1 | Morphological characterization of films via SEM and AFM. (a)** Characterization of the K-PHI film functionalized with F8BT (picture of a sample on the left) via SEM of a cross section. Secondary electron detector (right, top) and backscattered electron detector (ESB) (right, bottom) images are recorded, showing a rough surface with an average film thickness of ca. 1 µm. **(b)** AFM image of the film shown in (a). A root mean square roughness (RMS) of 35.8 nm is calculated. **(c)** Characterization of the PEDOT:PSS film via SEM of a cross section. Secondary electron detector (right, top) and backscattered electron detector (ESB) (right, bottom) images are recorded, showing a smooth surface with an average film thickness of ca. 0.6 µm. **(d)** AFM image of the film shown in (c). A root mean square roughness (RMS) of 2.59 nm is calculated.

# 2. Electrochemical study of HTM materials

The list of suitable hole transport materials (HTM) is vast, both for small molecule HTMs and conductive polymer HTMs. We chose multiple candidates based on their suitably positioned valence band potential and other desirable characteristics, such as conductivity, size or other functionalities (e.g., a band gap, which could allow them HTM to participate as active material). Conductive polymers (poly(3-hexylthiophen-2.5-diyl) (P3HT), Poly[4,8-bis(5-(2-ethylhexyl)thiophen-2-yl)benzo[1,2-b;4,5-b']dithiophene-2,6-diyl-alt-(4-(2-ethylhexyl)-3-fluorothieno[3,4-b]thiophene-)-2-carboxylate-2-6-diyl)] (PTB7TH), Poly(N-vinylcarbazole) (PVK), Poly(9,9-dioctylfluorene-alt-benzothiadiazole) (F8BT)) and small molecule (N,N'-Bis(3-methylphenyl)-N,N'-diphenylbenzidine (TPD), 1,3,5-Tris(N-carbazolyl)benzene (TCP), Tris-(4-carbazoyl-9-yl-phenyl)-amin (TCTA)) HTMs were chosen and deposited onto K-PHI films (on a transparent conductive substrate (here ITO) via dip coating) via spin coating. A solution of HTM in chloroform (1 mg mL$^{-1}$) was prepared, spin coated at 2,000 rpm for 30 s and subsequently annealed for 10 min at 80 °C on a hot plate.

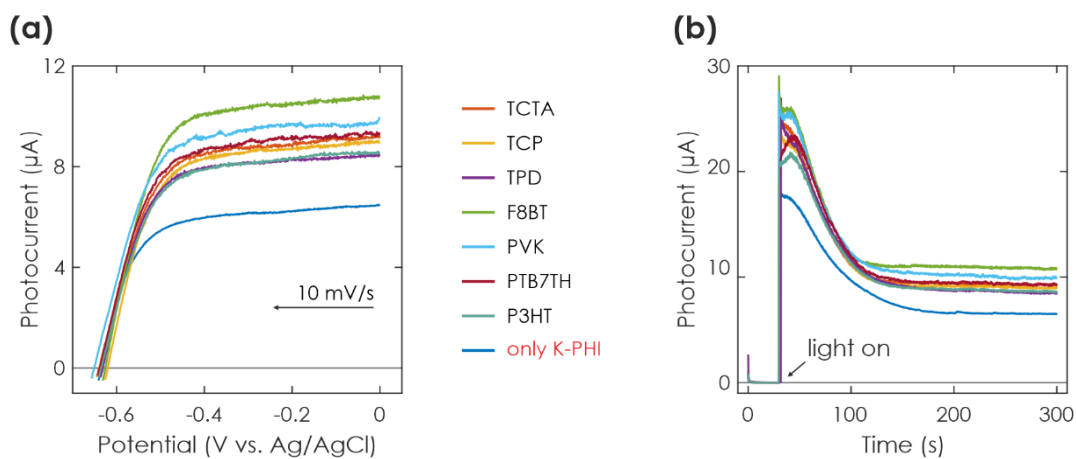

**Fig. S2.1 | Electrochemical characterization of K-PHI samples with different HTMs. (a)** LSV measurements of K-PHI with respective HTMs in a 3-electrode configuration and under illumination. All HTM materials show a larger photocurrent than the K-PHI sample alone, with F8BT showing the largest improvement. **(b)** Photocurrent output, when applying a bias voltage of 0 V to the device. An initial large photocurrent is observed, which stabilizes after ca. 100 s onto a "steady state" photocurrent.

To evaluate the performance of HTM materials, the photocurrent at different potentials was evaluated in a 3-electrode CV measurement, in the presence of a sacrificial electron donor to replace the HSM and under illumination with 1 sun. In Fig. 5 (a), the IV curves of the sample with different HTMs are shown. The potential is swept into the negative direction at a scan rate of 10 mV/s. From 0 to about -0.4 V vs Ag/AgCl, the photocurrent is very slowly decreasing due to a reducing driving force of the electron extraction from K-PHI. Subsequently, an open circuit potential (OCP) of approx. -0.65 V vs Ag/AgCl can be measured for all samples. F8BT shows the largest photocurrents and has thus been chosen for the further development of the device.

## 3. Activation & reset of solar battery samples

Prior to each solar battery measurement, an activation measurement was performed to ensure comparability between different samples. An exemplary measurement routine is shown in Fig. S3.1.

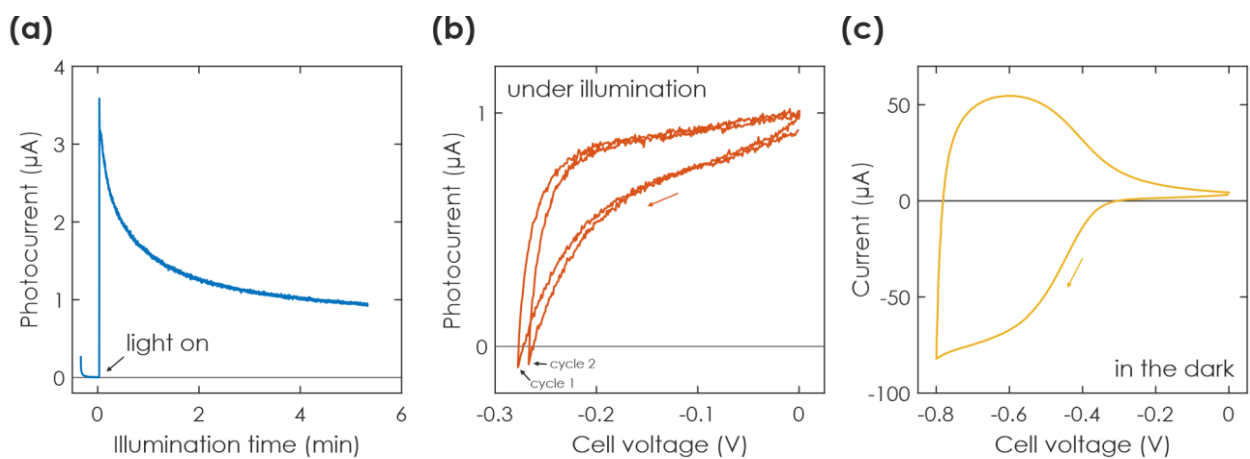

**Fig. S3.1 | Activation measurement routine. (a)** Photocurrent when applying a bias voltage of 0 V to the solar battery device. The bias was applied for 5 min. **(b)** CV under illumination and between 0 V and OCP, cycled twice with a scan rate of 10 mV/s. **(c)** CV in the dark, cycled once with a scan rate of 50 mV/s.

First, a bias voltage of 0 V was applied under illumination for 5 min. This ensures that possible organic contaminations are oxidized by K-PHI. Subsequently, a CV was measured under illumination between 0 V and the OCP (i.e., as soon as a cathodic current is observed). The idea behind this measurement is to control whether the sample shows the typical IV curve and is not short circuited (e.g., through accidental pinhole formation during sample preparation). Finally another CV measurement was carried out in the dark between 0 V and -0.8 V to check that both K-PHI and PEDOT:PSS work as intended when charged electrically in the dark.

Subsequently, a reset measurement was performed to remove all charges left on the device, i.e., electrons from K-PHI and holes from PEDOT:PSS, and to ensure that the sample was in a state that is comparable to other samples at different stages of the measurement. This measurement was performed both after the activation discussed in the paragraph above and after GCD measurements. An exemplary measurement is shown in Fig. S3.2.

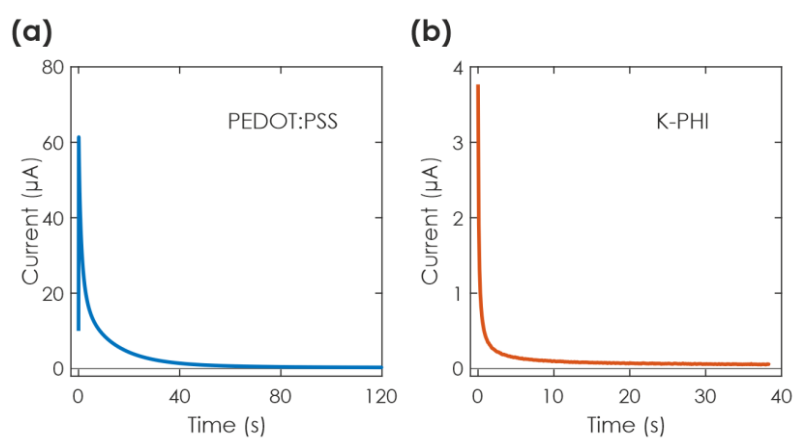

**Fig. 3.2 | Reset measurement routine.** An exemplary reset measurement is shown for PEDOT:PSS **(a)** and subsequently K-PHI **(b)**. A bias potential was applied to the respective electrodes of the device and a bias potential of -0.1 V vs. Ag/AgCl (PEDOT:PSS) and 0 V vs. Ag/AgCl (K-PHI) was applied. The measurement was stopped as soon as the current reached a value < 50 nA.

The concept behind this measurement is to remove unwanted charges from the electrodes individually, i.e., first remove holes from PEDOT:PSS, then electrons from K-PHI. It ensures that the charge state (or "oxidation state") of K-PHI and PEDOT:PSS is in balance prior to each measurement. For this, only one electrode was connected as working electrode, and a bias potential of 0 V vs. Ag/AgCl was applied against a reference electrode and a counter electrode which were immersed in the same electrolyte as the sample. The potential was applied until the current reached a value of < 50 nA. This measurement was first done for PEDOT:PSS and then for K-PHI.

# 4. Kinetic analysis of charge storage mechanism

In this section, we present a more detailed kinetic analysis of the charge storage mechanism of the two half cells (i.e., K-PHI and PEDOT:PSS; measured in a 3-electrode setup in a degassed 0.1 M KCl electrolyte against an Ag/AgCl reference electrode and Au counter electrode) as well as the full cell via a scan rate dependent CV analysis. Results are shown in Fig. S4.1.

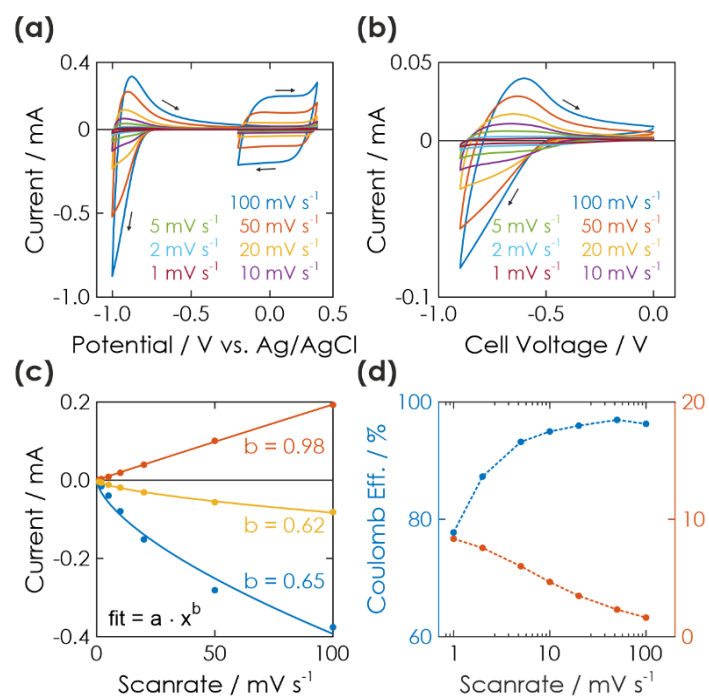

**Fig. S4.1 | Kinetic study via CV and with different scan rates. (a)** Study of K-PHI between 0.0 and -1.0 V vs. Ag/AgCl, and PEDOT:PSS between 0.5 and -0.1 V vs. Ag/AgCl for different scan rates. **(b)** Study of a full cell solar battery sample between 0 and -0.9 V vs. Ag/AgCl for different scan rates. **(c)** Extracted current of half cells ((a): K-PHI in blue, extracted at -0.9 V vs. Ag/AgCl; PEDOT:PSS in red, extracted at 0 V vs. Ag/AgCl) and full cell ((b), at -0.9 V, yellow) in dependence of the scan rate. A fit was calculated using the formula in the inset. [1] **(d)** Extracted charge and coulombic efficiencies from CV full cell measurements shown in (b). Normalized onto mass of K-PHI, HTM, and PEDOT:PSS

We will first discuss the half cells. K-PHI shows its very specific CV shape (Fig. S4.1 (a), the curve at more negative potentials). When extracting the current at the anodic sweep peak and plotting it against the scan rate, the data can be fitted according to the equation in the inset of Fig. S4.1 (c). The exponent b of 0.65 unveils largely faradaic or pseudocapacitive charge storage kinetics, [1,2] in line with our previously reported results. [3] On the other hand, when extracting current of the PEDOT:PSS measurement at a potential of 0 V vs. Ag/AgCl, the fit reveals a much more linear dependence of current with scan rate (b value of 0.98). Since this is more reminiscent of a capacitive kinetic signature, [4] but the charge storage mechanism of PEDOT:PSS is known to be faradaic, we link the kinetic behavior to pseudocapacitve charge storage, [1,2,5,6] as widely reported for PEDOT:PSS. [7-10]

Next, when looking at the scan rate dependence of peak current of the full cell (Fig. S4.1 (b)) and fitting the data (Fig. S4.1 (c)), a b value of 0.62 can be calculated, indicating a faradaic or pseudocapacitive charge storage mechanism. Note that this value is similar to the K-PHI half-cell, suggesting that the kinetics of charge storage of the full cell is dominated by K-PHI photoanode, which thus also limits the performance of the device. When looking at the extracted charge (Fig. S4.1 (d)) as a function of scan rate, we observe that with slower scan rates more charge can be extracted. This indicates that the amount of charge is limited by the charging kinetics, i.e., a very slow charging and discharging process maximizes the effective capacity – analogous to GCD measurements discussed in the main text in Fig. 3 (d). When looking at the coulombic efficiency as a function of scan rate of the full cell (Fig. S4.1 (d)), we see a maximum of 90 % at a scan rate of 50 mV/s. For both larger and smaller scan rates the coulombic efficiency decreases. We explain this with the existence of an optimum scan rate determined by kinetic limitations for larger scan rates and increased self-discharge for smaller scan rates, as previously reported for the K-PHI half cell. [3]

# 5. Additional analysis: Light charging and electric discharging profiles

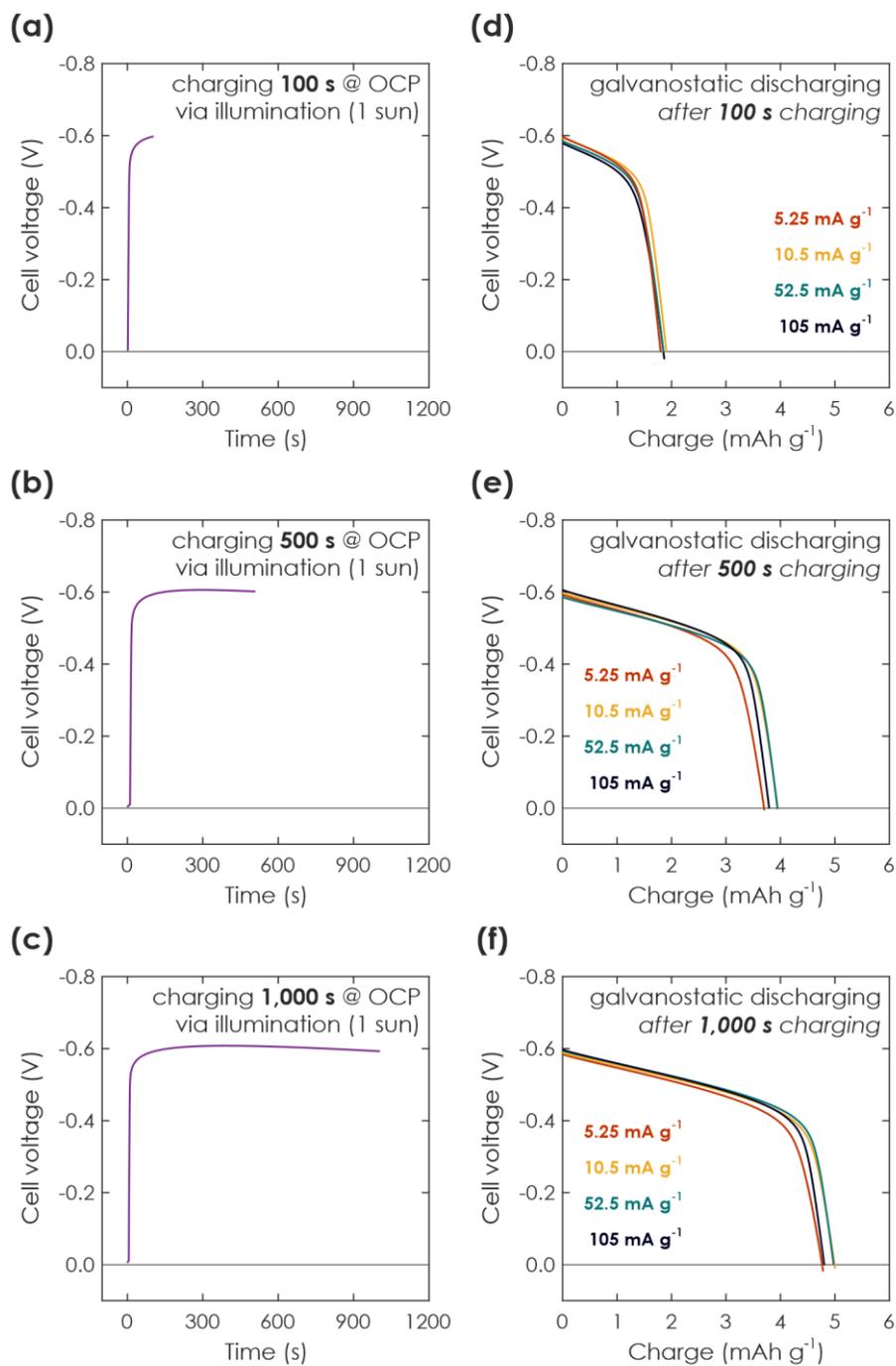

**Fig. S5.1 | Kinetic study of light charging and subsequent immediate electric discharging, discussed in the main text in Section 2.2.2. (a-c)** Charging of a solar battery sample via 1 sun illumination for three different illumination times. The solar battery sample is kept under OCP conditions to prevent any electric current during the light charging process. A photopotential of -0.6 V develops for all illumination times. **(d-f)** Respective electric discharging in the dark and with different currents given in the legend.

# 6. Additional analysis: Scaling of Power and Energy with Discharge Current

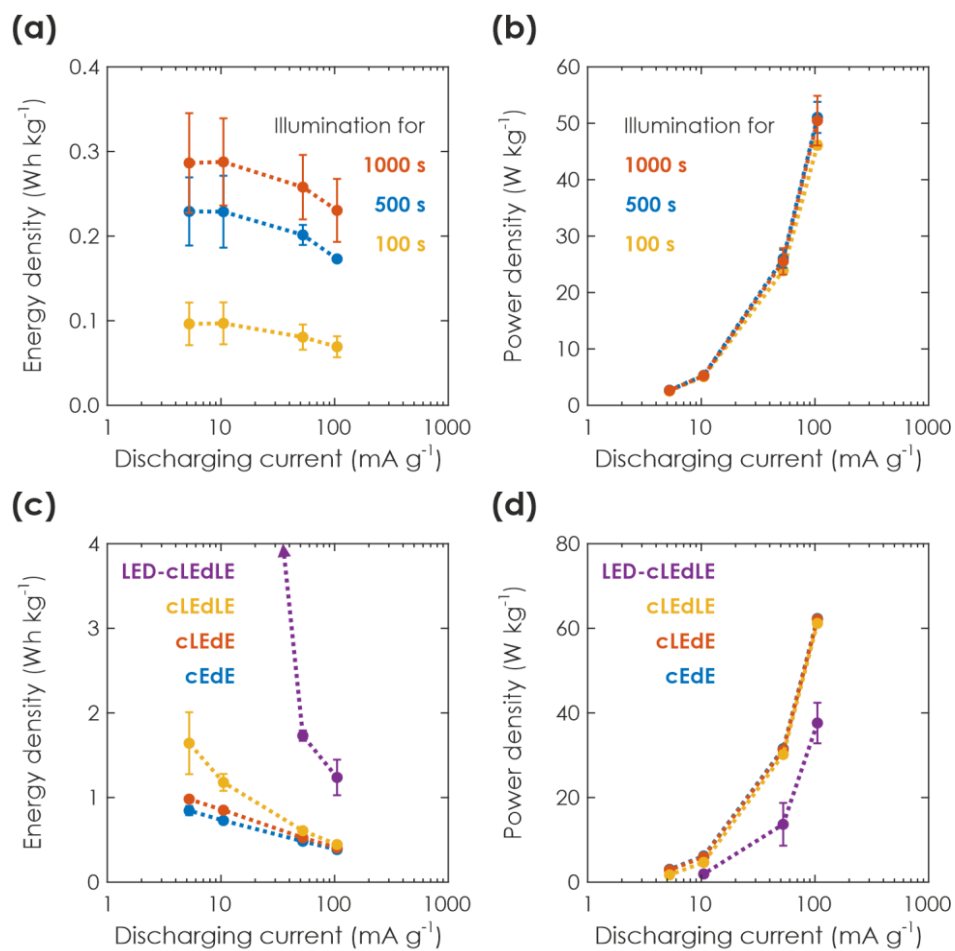

**Fig. S6.1 | Additional analysis to main text Fig. 2 & 3. (a-b)** Charging of a solar battery sample via 1 sun illumination and subsequent electric discharging in the dark at different discharging currents. Scaling of energy (a) and average power (b) with discharging current. The corresponding Ragone plot is shown in the main text in Fig. 2 (f). **(c-d)** Charging of a solar battery sample via GCD in the dark (cEdE), charging under illumination (cLEdE) or both charging and discharging under illumination (cLEdLE when illuminated with the solar simulator and 1 sun (100 mW cm$^{-2}$), or LED-cLEdLE when illuminated with an LED at 365 nm (100 mW cm$^{-2}$)). Scaling of energy (a) and Power (b) with discharging current. The corresponding Ragone plot is shown in the main text in Fig. 3 (f).

# 7. Shape of GCD measurements & chosen voltage window

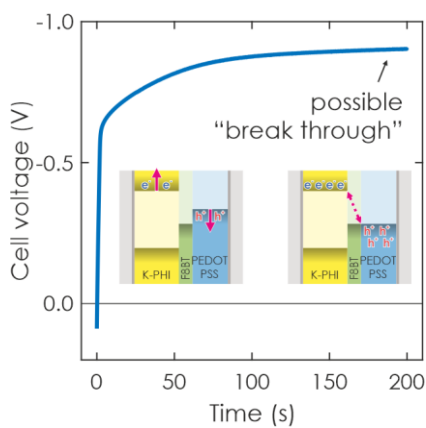

**Fig. S7.1 | Analysis of shape of GCD curve during charging.** GCD measured under analogous conditions as electric and light assisted electric GCD performance analysis discussed in the main text in Section 2.2.3 and shown in Fig. 3. The only difference is that the abort condition of charging is not the upper voltage vertex of 0.8 V. Instead, the measurement is stopped after 200 s. Inset shows normal charging (left) and a proposed short circuit or "breakthrough" mechanism (right).

In the following we will discuss the rationale behind the chosen GCD voltage window. The shape of the charging curve for a typical battery shows an initial fast increase to a relatively flat plateau in which the battery experiences charging. Subsequently after reaching a fully charged state, the potential increases rapidly again. [1,4] Conversely, in our solar battery device we do not observe this second rapid increase. Instead, the plateau seems to flatten out (Fig. S7.1). We propose a possible mechanism for this behavior in the inset in Fig. S7.1: Electrons from K-PHI can reach holes from PEDOT:PSS, a process that is normally inhibited by the HTM F8BT and termed by us as "breakthrough" voltage. This is possible as soon as the potential of hole storage in PEDOT:PSS (i.e., the valence band of PEDOT:PSS) gets more negative than the valence band of F8BT. Consequently, holes are being injected from PEDOT:PSS into F8BT and can then recombine with electrons from K-PHI.

Note also that while photocharging of the solar battery occurs at a potential of 0.6 V (see main text Fig. 2 (b) and (c)), we reach a higher voltage of 0.8 V before the "break through" voltage is reached. We explain this by the different charging mechanisms of light charging via an "internal" photocurrent and electric charging via applying the "external" charging current. While the first charges the K-PHI film most likely more in the bulk and close to the junction to the HTM, the latter charges the film via the substrate. The low conductivity of K-PHI produces an iR drop across the film, which in return reduces the voltage at which the device is charged and subsequently discharged. This means: K-PHI charging starts at 0.6 V, and charges can be theoretically stored on the device until it reaches the "break through" voltage. However, while electric charging with a very small current should in theory stay at a similar potential than light charging and yield a similar capacity (note the bend in the cell voltage at ca. 0.6 V, when charged electrically as shown in Fig. S7.1), this current would produce a much smaller power output. Hence, we chose 0.8 V as a compromise for electric charging, which – while charging the device unevenly (i.e., the parts of K-PHI close to the substrate are charged more than the bulk) – produces a reasonably high power and energy output at currents between 5.25 and 105 mA g$^{-1}$. Note that in theory the capacity of K-PHI is thus underestimated in this work.

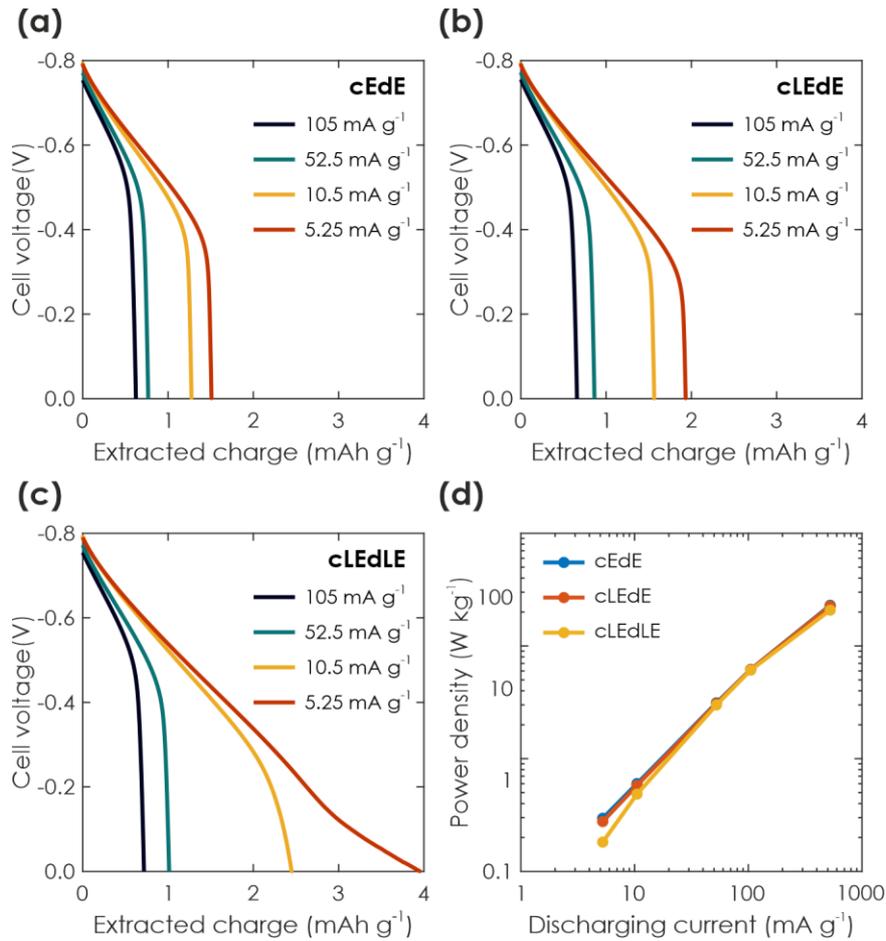

**Fig. S7.2 | Analysis of shape of GCD curve during discharging. (a-c)** GCD discharge profiles with different discharge currents and in the dark (cEdE) (a), illuminated during only charging (cLEdE) (b) or illuminated during both charging and discharging (cLEdLE) (c), measured in analogous conditions than measurements shown in the main text in Fig. 3 (d), (e), (f). **(d)** Power density of solar battery samples when GCD is performed in cEdE (blue; extracted from (a); hidden behind red curve), cLEdE (red; extracted from (b)), or cLEdLE (yellow; extracted from (c)). Power output reproduced from the main text Fig. 3 (f).

Next, we will address the shape of GCD discharge curves under different illumination conditions as shown in Fig. S7.2 (a-c). Conditions are analogous to measurements shown for GCD in the main text in Fig. 3 (d)-(f). Note that for cLEdLE and slower currents after discharging the plateau, the cell voltage does not rapidly decrease to zero but rather shows a tail. This is due to the photocurrent generated during illumination which simultaneously charges the device during electric discharging. It affects small electric discharging currents more since they are in the range of the photocurrent, as discussed in the main text. The power density output is thus affected and decreases for small discharging currents when discharged under illumination ( Fig. S7.2 (d)).

# 8. Charge retention for delayed discharge

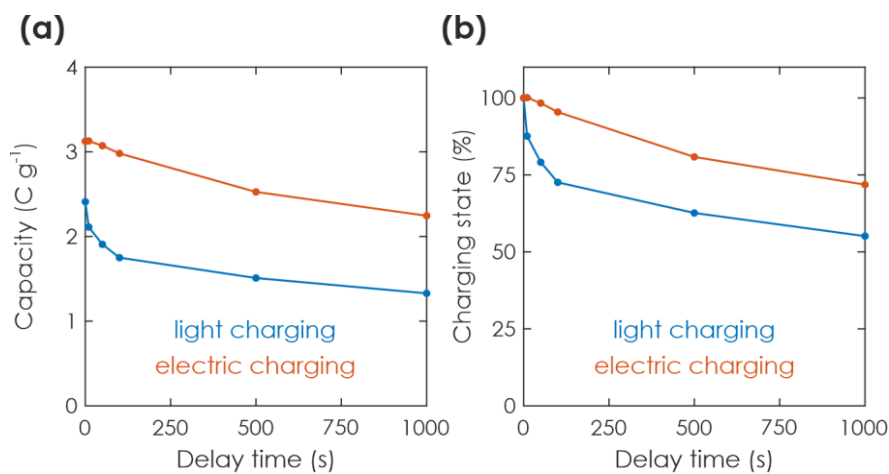

**Fig. S8.1 | Analysis of charge retention.** (a) Extracted charge from a solar battery, which was charged via illumination at OCP and discharged in the dark with 10.5 mA g$^{-1}$ (blue), or charged and discharged in the dark (between a cell voltage of 0 V (discharged) and 0.8 V (charged)) with 10.5 mA g$^{-1}$ (red). The subsequent electric discharge in the dark (10.5 mA g$^{-1}$) was delayed as shown in the Fig.. (b) Extracted charge retention from (a), normalized against the extracted charge without any time delay before the discharge.

An important feature of solar battery device is its charge retention time, i.e., how fast the solar battery self-discharges. To analyze this, we perform a measurement analogous to light charging and electric discharging measurements discussed in the main text in Section 2.2.2, but fix the illumination time onto 1000 s and instead delay dark discharging (10.5 mA g$^{-1}$). The resulting charge is shown in Fig. S8.1 (a) blue data points. The charging state describing how much charge is left after the delayed discharging is shown in Fig. S8.1 (b). We observe an initial fast decay of the charging state which subsequently levels out. After 500 s and 1000 s, a respective 63 % and 55 % of the initial charge was left on the device. We also analyze delayed discharge when charged not with light but with an electric charging current (10.5 mA g$^{-1}$) in the dark (Fig. S8.1 (a) and (b) red data points). Interestingly, while under the chosen conditions a larger initial charge was present on the device (3.12 C g$^{-1}$ for electric charging vs. 2.41 C g$^{-1}$ for light charging), a delayed discharge decreases the charge output less than for light charging (after 500 s and 1000 s, a respective 81 % and 72 % of the initial charge was left on the device).

We rationalize the self-discharge with issues arising from sample preparation, such as pinholes, or more generally, insufficiently charge selective HTM performance to separate charges from the cathode and anode efficiently. Electrons and holes can thus slowly recombine via the internal layer, and discharge the system from the inside, i.e. by the electrode material adjacent to the HTM. The faster self discharge from light charging can be explained with light charging occurring more in the bulk of K-PHI and electric charging occurring more close to the junction to the substrate. Since the former charges K-PHI closer to the junction of the HTM and due to the low conductivity of K-PHI, light charging produces a slightly faster self discharge. In the electrically charged case, however, the material adjacent to the contact is charged first, thus hindering charge recombination through the material in the device volume.

# 9. Solar cell performance comparison 1 sun & 365 nm LED

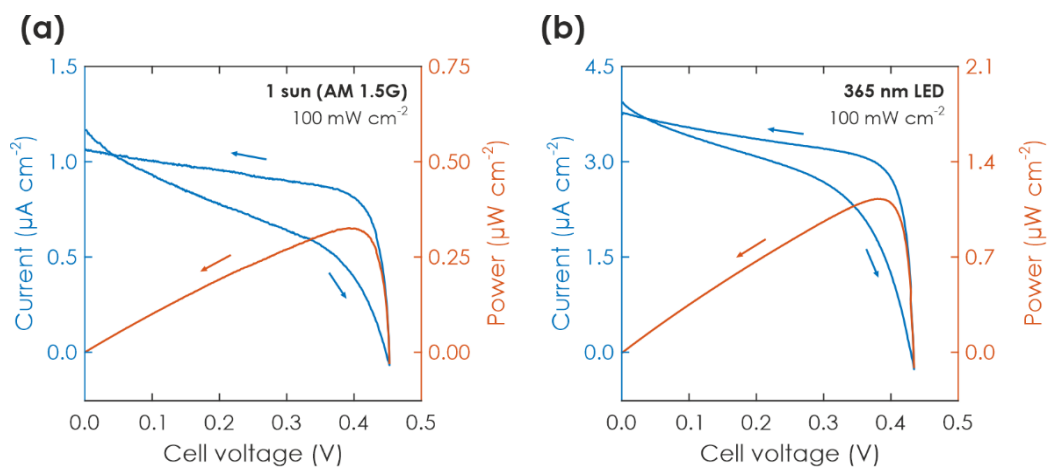

**Fig. S9.1 | Comparison of IV curves with different light sources. (a)** Current-voltage (blue) and power (red) curves (10 mV s$^{-1}$) of a solar battery sample in solar cell mode, illuminated with 1 sun. Reproduced from main text Fig. 2 (a). **(b)** Current-voltage (blue) and power (red) curves (10 mV s$^{-1}$) of a solar battery sample in solar cell mode, illuminated with a 365 nm LED (100 mW cm$^{-2}$; bandwidth of ca. 10 nm).

We discuss operation of the solar battery device in solar cell mode extensively in main text Section 2.2.1. Two different illumination sources are utilized for measurements presented in this work: Either artificial sunlight (1 sun) with AM 1.5 global standard (see Methods section for details) to provide real-world illumination conditions for operation of the solar battery, or an 365 nm LED (ca. 360-375 nm) to provide light which is below the bandgap of K-PHI (450 nm). Both light sources were operated at the same power setting (100 mW cm$^{-2}$). IV and power curves are shown in Fig. S9.1 and performance parameters are summarized in Table S9.1. Notably, the shape of IV and power curves looks very alike and open-circuit potential, potential at which maximum power is provided as well as the fill factor are very similar. However, as expected short-circuit current is significantly higher when operated with the LED (277 %) and the device produces much more power (248 %).

**Table S9.1 | Performance parameter of IV curves. (a)** Summary of performance parameters of current-voltage and power curves of the solar battery, when illuminated either with artificial sunlight or a 365 nm LED. The fill factor (FF) is calculated via the backward sweep (OCP to 0 V) from open circuit potential ($U_{OCP}$), short circuit current ($I_{SC}$), and potential ($U_{Pmax}$) as well as current ($I_{Pmax}$) at which the device produces maximum power ($P_{max}$). IV curves are shown in Fig. S9.1.

| Illumination mode | $U_{OCP}$ (V) | $U_{Pmax}$ (V) | $I_{sc}$ (µA cm$^{-2}$) | $I_{Pmax}$ (µA cm$^{-2}$) | $P_{max}$ (µW cm$^{-2}$) | FF |
|---|---|---|---|---|---|---|
| 1 sun (100 mW cm$^{-2}$) | 0.45 | 0.39 | 1.07 | 0.828 | 0.326 | 0.68 |
| 365 nm LED (100 mW cm$^{-2}$) | 0.43 | 0.38 | 3.78 | 2.98 | 1.13 | 0.69 |